# Pump-probe model for the Kramers-Kronig relations in a laser

H. Yum[1], M. S. Shahriar[1,2]

[1] *Department of Electrical Engineering and Computer Science, Northwestern University, Evanston, IL 60208*

[2] *Department of Physics and Astronomy, Northwestern University, Evanston, IL 60208*

**Abstract:**
In this paper, we study theoretically a pump-probe model for the Kramers-Kronig (KK) relations during laser operation. A laser gain medium at steady state becomes saturated and the lasing field experiences a flat gain equal to the cavity loss. A solution of the laser equations reveals that the lasing frequency experiences a dispersion that is linear over the allowed bandwidth. However, outside this band, the lasing stops, so that the dispersion is that of the unsaturated gain medium. The combined profile is therefore non-analytical, and cannot be explained in terms of the KK relations. In order to interpret this situation, it is important to consider carefully the physical basis of the KK relations and its connection to causality. We conclude that the KK relation is expected to apply only to an independent probe applied to the medium, which is under excitation by the pump producing the gain as well as the lasing mode. The absorption/gain and dispersion profiles are then analytical, and satisfy the KK-relations. Specifically, these are variants of the so-called Mollow-Ezekiel spectra of probe absorption/gain and dispersion in the presence of a pump, with the exception that in this case the medium is inverted.

PACS numbers: 42.50.Gy, 42.55.Ah, 42.60.Da

Dispersion in a laser cavity is an important issue in high precision laser interferometric measurements such as rotation sensing, vibrometry and gravitational wave detection[1]. For these applications, the sensitivity of the cavity is defined as the ratio of the amount of a resonance frequency shift to a particular change in length of the cavity, and is enhanced by tailoring dispersion[1,2]. Recently, we have shown[1] that by using a gain profile with a dip at the center, it is indeed possible to realize a so-called superluminal laser with the property that the group velocity becomes much larger than the free space speed of light (without violating special relativity), and its frequency is very sensitive to a change in the length of the cavity. While this result follows from a careful analysis of the laser equations[3], the resulting behavior of the real and imaginary parts of the effective susceptibility for the lasing mode is far from obvious. In fact, we find that both the saturated gain and the dispersion profiles become non-analytical at the frequencies where the unsaturated gain equals the cavity loss. As such, it is impossible to interpret these results using the Kramers-Kronig (KK) relations[4,5,6,7], or even the modified KK (MKK)[8,9,10] relations that apply to a system excited by a saturating probe.

This non-analyticity is present even for a simpler system where the unsaturated gain profile is simply Lorentzian, without a dip. Therefore, such a behavior is a generic property of a single mode laser. Of course, since the underlying equations are causal, the non-analytic behavior of the effective gain and dispersion must also be causal. Thus, one might expect the KK relations to apply, since these relations are dictated by requiring simply that the response of a system be causal. In this paper, we explain why the KK relations do not hold. Specifically, we point out, as explained below, that the problem lies with defining what the probe really is, in the context of the KK relations. We then develop a model, using an external probe, for a gain and dispersion that become analytic and do satisfy the KK relations.

In order to understand why the KK relations do not apply in the conventional sense, it is instructive first to the recall the basic argument underlying these relations[5]. Consider a system excited by a probe pulse of a finite duration. By definition, such a pulse has a spectral distribution. Each component of this spectrum is of infinite extent in time, covering the past and the future. The interference between these components determines the shape of the pulse in general, and the front edge in particular. The medium will absorb/amplify each component by a different amount, and shift its phase by a different amount. These two processes must depend on in each other in a way so that the front edge of the pulse does not pass through the medium at a speed exceeding the vacuum speed of light. This requirement leads immediately to the KK relations.

It is also instructive to consider the mathematical foundation for the KK relations[7]. The response function $\chi(t-t')$ describes how a system responds to a perturbation. This function must be zero for $t<t'$, since a system cannot respond to a perturbation before it is applied. This causality condition implies that the Fourier transform of the response, $\chi(\omega)$, is analytic in the upper half of the complex plane[7]. Furthermore, for a real system $\chi(\omega)$ must vanish as $\omega$ becomes very large. These conditions immediately yield the KK relations in the form of Hilbert transforms.

Consider now a single mode laser in steady state. Suppose we treat the gain medium to be the system of interest. The lasing field then cannot be treated as a perturbation of the type discussed above, since the amplitude as well as the phase of this field are not decoupled from the system. Thus, the effective gain and dispersion profiles of the lasing mode are not expected to satisfy the KK relations. Therefore, the knowledge of the absorption/gain profile under lasing condition would not allow one to infer the corresponding effective dispersion profile by applying the KK relations. In reference 10, the authors attempt to determine analytically the dispersion profile of a laser based on the gain profile determined by solving the rate equations. We believe this to be inaccurate for several reasons. First, the rate equations ignore important coherence effects that nullify this approach. The authors of ref. 10 claim that the result should still be valid in some limit, without proper justification. Second, and more importantly, they do not present any argument as to why the KK-relations should apply, in the context of the argument we presented above. Finally, to our knowledge, there is no experimental evidence validating the results presented in ref. 10.

The only sensible way to analyze the causality of the lasing system would be to apply an independent perturber, such as a separate probe field whose input amplitude, central frequency and temporal profile can be varied without affecting the system. In this paper, we consider such a probe, and determine the real and imaginary parts of the susceptibility of the lasing system. The resulting spectral profiles are analytic over all frequencies, and satisfy the KK relations readily.

Before presenting this analysis, we first analyze the effective susceptibility of a lasing field in steady state. It is instructive to consider the imaginary part $\chi''$ and real part $\chi'$ of the effective susceptibility in two separate domains: the spectral range where the system is lasing, and the range over which no lasing occurs. For the gain medium, we consider an inverted two level system. The susceptibility of such a system can be expressed as[1]

$$\chi'' = -G\left(\frac{\Gamma^2}{2\Omega^2 + \Gamma^2 + 4(\nu-\nu_0)^2}\right) \tag{1.1}$$

$$\chi' = G\frac{2(\nu-\nu_0)}{\Gamma}\left(\frac{\Gamma^2}{2\Omega^2 + \Gamma^2 + 4(\nu-\nu_0)^2}\right) \tag{1.2}$$

Here $\Gamma$ is the linewidth of the two level system, $\Omega$ is the Rabi frequency equal to $\wp E/\hbar$, where $\wp$ is the dipole moment. Using the Wigner–Weisskopf model[11] for spontaneous emission we can define two parameters: $\xi \equiv \wp^2/(\hbar^2\Gamma)$. In terms of this parameter, the Rabi frequencies can be expressed as $\Omega^2 = \Gamma E^2 \xi$, and the gain parameter can be written as $G \equiv \hbar N \xi/\varepsilon_0$, where $\varepsilon_0$ is the permittivity of free space, and N represents the density of the two level systems.

In what follows, we make use of the field amplitude equation of the semi-classical laser theory[3]. First, we consider light in a ring laser cavity which contains a gain medium. The amplitude of the field is described by:

$$\dot{E} = -\frac{1}{2}\frac{\nu}{Q}E - \frac{1}{2}\nu E \chi''(E,\nu) \quad (2)$$

where $\nu$ is the lasing frequency, $E$ is the laser field amplitude, and Q is the cavity quality factor. We assume $\nu_0$ to be the frequency around which $\chi''$ is symmetric. We first solve Eq. (2) in steady state ($\dot{E}=0$), so that $\chi''(E,\nu) = -1/Q$ for $E \neq 0$. Hence, in the lasing range, the gain is saturated to be equal to the cavity loss. Using Eq. (1.1) and $\chi'' = -1/Q$, we obtain the lasing frequency $\nu = \nu_0 \pm \sqrt{(QG\Gamma^2 - 2\Omega^2 - \Gamma^2)/4}$. From this expression, we find $\Omega^2 = QG\Gamma^2 - \Gamma^2 - 4(\nu - \nu_0)^2$ which is proportional to the lasing field intensity, and is a function of the lasing frequency. The boundaries $\nu_1$ and $\nu_2$ of the lasing range are found by setting $\Omega=0$: $\nu_{1,2} = \nu_0 \pm \sqrt{(QG\Gamma^2 - \Gamma^2)/4}$. For $\nu_1 < \nu < \nu_2$, we have $\chi'' = -1/Q$. From Eq.(1) we can derive the ratio $\chi'/\chi'' = -2(\nu - \nu_0)/\Gamma$. Substituting $\chi'' = -1/Q$ into $\chi'/\chi''$, we get $\chi' = 2(\nu - \nu_0)/(Q\Gamma_e)$. For $\nu < \nu_1$, $\nu > \nu_2$, the gain cannot compensate for cavity loss so that $\Omega=0$. Therefore, in this range, we get $\chi'' = -G\Gamma^2/[\Gamma^2 + 4(\nu - \nu_0)^2]$ and $\chi' = [2(\nu - \nu_0)/\Gamma]\left(G\Gamma^2/[\Gamma^2 + 4(\nu - \nu_0)^2]\right)$. To summarize:

$$\text{For } \nu_1 < \nu < \nu_2,\ \chi'' = -\frac{1}{Q},\ \chi' = \frac{2(\nu - \nu_0)}{Q\Gamma},\ \Omega^2 = QG\Gamma^2 - \Gamma^2 - 4(\nu - \nu_0)^2 \quad (3.1)$$

$$\text{For } \nu < \nu_1 \text{ or } \nu > \nu_2,\ \chi'' = -\frac{G\Gamma^2}{\Gamma^2 + 4(\nu - \nu_0)^2},\ \chi' = G\frac{2(\nu - \nu_0)}{\Gamma}\left(\frac{\Gamma^2}{\Gamma^2 + 4(\nu - \nu_0)^2}\right),\ \Omega^2 = 0 \quad (3.2)$$

Fig.1 graphically illustrates $\chi'$, $\chi''$ and $\Omega^2$ of Eq. (3.1&2). For illustration, we have chosen $\Gamma = 2\pi \times 10^9 s^{-1}$, $G = 3 \times 1/Q$. Note that according to Eq. (2), $Q = \nu_0/\Gamma_c$, where $\Gamma_c$ is the empty cavity linewidth. We have chosen $Q=3.8 \times 10^8$ corresponding to $\Gamma_c = 2\pi \times 10^6 s^{-1}$ for $\nu_0 = 2\pi \times 3.8 \times 10^{14}$ corresponding to the $D_2$ transition in Rubidium atoms.

Because of the "kinks" present at the lasing boundaries, both $\chi'$ and $\chi''$ are non-analytical e.g., the first derivative will be discontinuous. This means that the gain and dispersion profiles are valid only as we approach these point from either side, but these points have to be excluded. Physically, this is due to the fact that the laser is at the threshold (gain just matching loss), a condition under which the semi-classical laser equations are not valid. This non-analyticity implies that we cannot even attempt to apply the KK-relations. We are then faced with the question of what would be the proper interpretation of the KK-relations for this system.

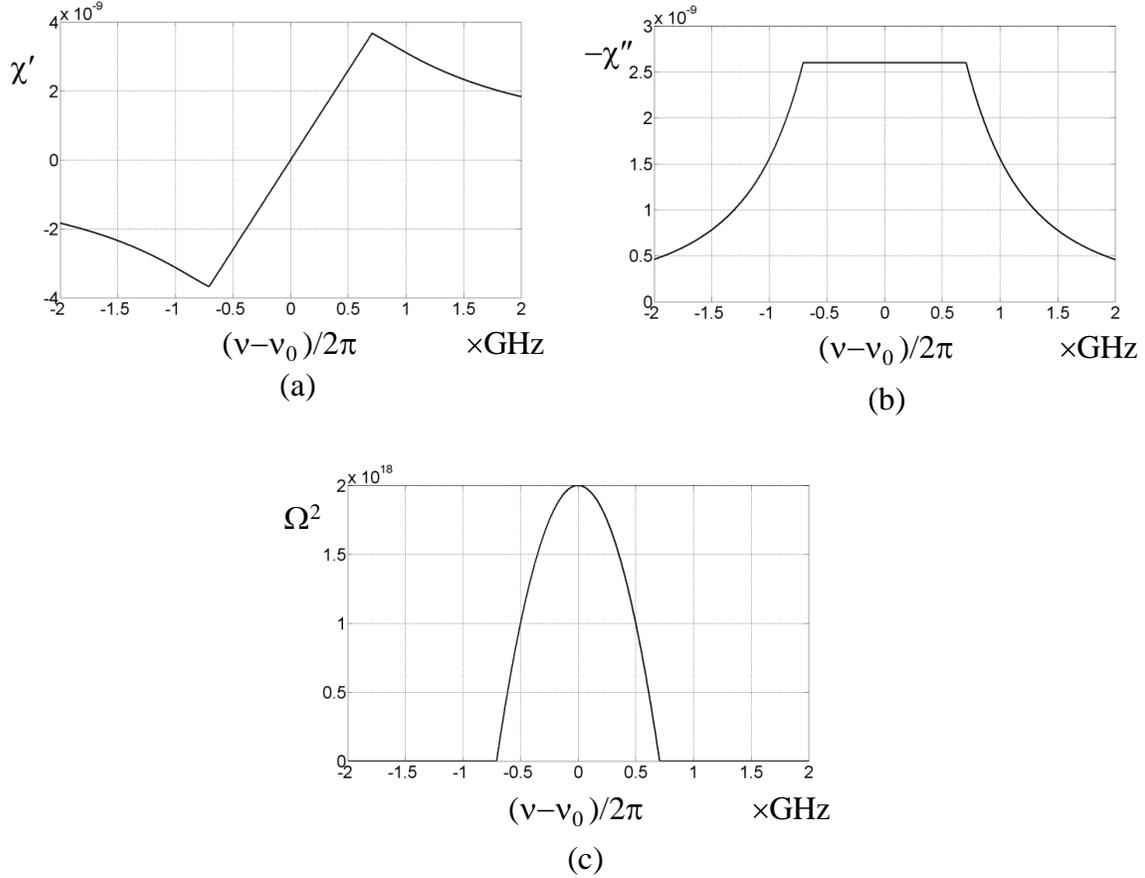

FIG.1 Illustration of effective gain, dispersion and intensity profiles for a ring laser under steady state operation (a) $\chi'$ (b) $-\chi''$ (c) $\Omega^2$

As discussed above, the answer to this question can be obtained by considering a separate, spectrally tunable probe applied to this medium. This is illustrated in Fig.2. The cavity loss due to the beam splitters would be incorporated into Q. We can now consider the response of the gain medium as the probe frequency is scanned. The absorption ($\chi''$) and dispersion ($\chi'$) experienced by the probe

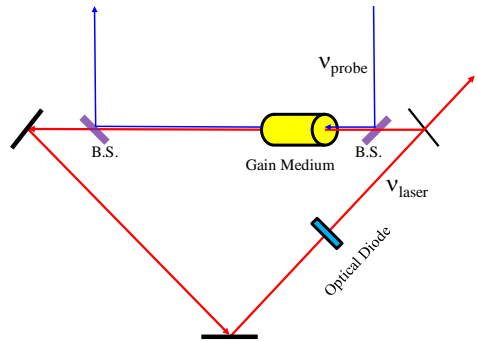

FIG.2 Schematic of a laser cavity in the presence of a probe which is introduced to observe the gain and dispersion of the medium. B.S.; Beam splitter. $\nu_{probe}$; Probe frequency, $\nu_{laser}$; Lasing frequency

would then satisfy the KK relations. For the spectral range ($\nu_{laser} < \nu_1$ or $\nu_{laser} > \nu_2$) where no lasing occurs, $\chi'$ and $\chi''$ would obviously be the same as those plotted in Fig.1, in the limit of a vanishingly small probe field. For the lasing range ($\nu_1 < \nu_{laser} < \nu_2$), $\chi'$ and $\chi''$ would depend on the values of $\nu_{laser}$ and $\Omega$. Qualitatively, these would be similar to what has been studied extensively in the context of probe gain produced by a driven two level atom [12,13,14,15,16,17]. However, an important difference here is that the two level system in consideration is also pumped incoherently in order to produce the population inversion.

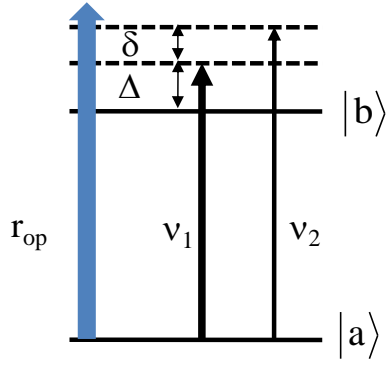

FIG. 3. Two level atom incoherently pumped for population inversion via optical pumping (Blue arrow) at the rate of $r_{op}$, and illuminated by a strong pump $\nu_1$ (the lasing field) and a weak probe $\nu_2$.

Fig. 3 illustrates schematically the physical process implied in Fig. 2. Specifically, we consider a collection of two level atoms which are incoherently pumped to induce population inversion. The atoms incoherently excited from the level $|a\rangle$ to $|b\rangle$ are illuminated simultaneously by a strong pump field $\nu_1$ (emulating the lasing mode) and a weak probe field $\nu_2$. The probe is weak in the sense that the atoms cannot be saturated by the probe alone. The interaction of the two coherent fields with the atom is described by the Hamiltonian:

$$H = -\Delta |b\rangle\langle b| - \frac{1}{2}\left[(\Omega_1 + \Omega_2 e^{-i\delta t})|b\rangle\langle a| + \text{h.c.}\right] \qquad (4)$$

Here, h.c. denotes Hermitian conjugate, $\Delta = \nu_1 - \omega_{ba}$ is the pump detuning from atomic resonance and $\delta = \nu_2 - \nu_1$ is the probe–pump detuning. $\Omega_1$ and $\Omega_2$ are the respective Rabi frequencies expressed as $\mu_{ba}E_k/\hbar$ in terms of the atomic–dipole matrix element $\mu_{ba}$ and the corresponding field amplitude $E_k$ (k=1 or 2). The equations of motion for the density matrix elements $\rho_{ij}$ (i,j= a or b) with phenomenological damping terms as well as the incoherent pumping terms are written as:

$$\dot{\rho}_{ba} = -(\gamma_{ba} - i\Delta + \tfrac{r_{op}}{2})\rho_{ba} - \tfrac{i}{2}(\Omega_1 + \Omega_2 e^{-i\delta t})(\rho_{bb} - \rho_{aa}) \qquad (5.1)$$

$$\dot{\rho}_{bb} = r_{op}\rho_{aa} - \Gamma\rho_{bb} + \tfrac{i}{2}(\Omega_1 + \Omega_2 e^{-i\delta t})\rho_{ab} - \tfrac{i}{2}(\Omega_1^* + \Omega_2^* e^{i\delta t})\rho_{ba} \qquad (5.2)$$

$$\dot{\rho}_{aa} = -r_{op}\rho_{aa} + \Gamma\rho_{bb} - \tfrac{i}{2}(\Omega_1 + \Omega_2 e^{-i\delta t})\rho_{ab} + \tfrac{i}{2}(\Omega_1^* + \Omega_2^* e^{i\delta t})\rho_{ba} \qquad (5.3)$$

Here $\Gamma$ and $\gamma_{ab}$ are the longitudinal and transverse decay rates, respectively, and $r_{op}$ is the incoherent optical pumping rate. We have used an interaction picture so that the fast oscillations of the off-diagonal elements at the frequency $\nu_1$ have been transformed out. We assume the system to be closed so that the trace of the density matrix is unity: $\rho_{bb} + \rho_{aa} = 1$. Using Eq. (5.2)

and (5.3), and the trace relation, we can readily derive the time evolution of the population inversion, $n \equiv \rho_{bb} - \rho_{aa}$:

$$\dot{n} = 2r_{op}\rho_{aa} - 2\Gamma n + i(\Omega_1 + \Omega_2 e^{-i\delta t})\rho_{ab} - i(\Omega_1^* + \Omega_2^* e^{i\delta t})\rho_{ba} \tag{6}$$

To find the steady state solution to Eq. (5.1) and (6), we assume that the weak probe perturbs the atomic dipole moment as well as the population inversion. Thus, in the interaction picture, $\rho_{ba}$ and n have main components as constants, with small components oscillating at $\pm\delta$ (ignoring higher order terms, which vanish in the limit of a very small probe amplitude). The solution to $\rho_{ba}$ is thus assumed to be of the form:

$$\rho_{ba} = \rho_{ba}^0 + \rho_{ba}^1 e^{-i\delta t} + \rho_{ba}^{-1} e^{i\delta t} \tag{7.1}$$

Likewise, n is written as

$$n = n^0 + n^1 e^{-i\delta t} + n^{-1} e^{i\delta t} \tag{7.2}$$

where $\rho_{ba}^0$ and $n^0$ are the solutions in the presence of only the pump field. $\rho_{ba}^{\pm 1}$ and $n^{\pm 1}$ are small complex amplitudes such that $|\rho_{ba}^{\pm 1}| \ll |\rho_{ba}^0|$, $|n^{\pm 1}| \ll |n^0|$. Our goal is to find $\rho_{ba}^1$ which is associated with the non-linear susceptibility of the atomic dipole oscillating at the probe frequency $\nu_2 = \nu_1 + \delta$. Inserting Eqs. (7) in Eqs. (5.1) and equating terms of order $e^{ik\delta t}$ (k=−1,0,1) on both sides of Eq.(5.1), we obtain

$$\dot{\rho}_{ba}^0 = -(\gamma_{ba} + \tfrac{r_{op}}{2} - i\Delta)\rho_{ba}^0 - \tfrac{i}{2}\Omega_1 n_0 \tag{8.1}$$

$$\dot{\rho}_{ba}^1 = -[\gamma_{ba} + \tfrac{r_{op}}{2} - i(\Delta + \delta)]\rho_{ba}^1 - \tfrac{i}{2}(\Omega_1 n^1 + \Omega_2 n^0) \tag{8.2}$$

$$\dot{\rho}_{ba}^{-1} = -[\gamma_{ba} + \tfrac{r_{op}}{2} - i(\Delta - \delta)]\rho_{ba}^{-1} - \tfrac{i}{2}\Omega_1 n^{-1} \tag{8.3}$$

In the same manner, we use Eqs.(6) and (7) to get:

$$\dot{n}^0 = r_{op} - \Gamma - (r_{op} + \Gamma)n^0 + i\Omega_1 \rho_{ba}^0 - i\Omega_1^* \rho_{ba}^0 \tag{8.4}$$

$$\dot{n}^1 = -(\Gamma + r_{op} - i\delta)n^1 + i\Omega_1 \rho_{ab}^{-1} - i\Omega_1^* \rho_{ba}^1 + i\Omega_2 \rho_{ab}^0 \tag{8.5}$$

$$\dot{n}^{-1} = -(\Gamma + r_{op} + i\delta)n^{-1} + i\Omega_1 \rho_{ab}^1 - i\Omega_1^* \rho_{ba}^{-1} - i\Omega_2^* \rho_{ba}^0 \tag{8.6}$$

Here, we have dropped terms which are the products of $\Omega_1$ and the small perturbed components such as $n^{\pm 1}, \rho_{ba}^{\pm 1}$. By solving Eqs. (8) in the steady state approximation, we get

$$n^0 = \frac{r_{op} - \Gamma}{\theta + |\Omega_1|^2 \eta / [\Delta^2 + \eta^2]} \tag{9.1}$$

$$\rho_{ba}^1 = \frac{\tfrac{1}{2}\Omega_2}{\Delta + \delta + i\eta}\left[1 - \frac{\frac{|\Omega_1|^2(\Delta - \delta + i\eta)(\delta + 2i\eta)}{2}\frac{1}{\Delta - i\eta}}{(\delta + i\theta)(\Delta + \delta + i\eta)(\delta - \Delta + i\eta) - |\Omega_1|^2(\delta + i\eta)}\right]n^0 \tag{9.2}$$

where $\eta = \gamma_{ba} + r_{op}/2$, $\theta = r_{op} + \Gamma$. In Eq. (9.1), note that $n^0=1$ for $r_{op} \gg \Gamma$, $\Omega_1=0$. In this limit, the atoms are population-inverted before the simultaneous presence of the pump and the probe fields.

To derive the susceptibility, we use the relation $\epsilon_0 \chi E_2 = 2N\mu_{ba}\rho_{ba}^1$, where N is the number of atoms, and $\epsilon_0$ is the vacuum permittivity. It leads us to an analytic expression of $\chi$ in the form of

$$\chi = G \frac{n^0 \gamma_{ba}}{\Delta+\delta+i\eta} \left[ 1 - \frac{\frac{|\Omega_1|^2 (\Delta-\delta+i\eta)(\delta+2i\eta)}{2 \quad \Delta-i\eta}}{(\delta+i\theta)(\Delta+\delta+i\eta)(\delta-\Delta+i\eta)-|\Omega_1|^2(\delta+i\eta)} \right] \quad (10)$$

where $G = N|\mu_{ba}|^2/(\epsilon_0 \hbar \gamma_{ba})$. It is instructive to plot the real ($\chi'$) and imaginary ($\chi''$) parts of this susceptibility, as shown in figure 4. For illustration, we choose $\Gamma=2\pi\times10^7 s^{-1}$, $\gamma_{ba}=2\pi\times5\times10^6 s^{-1}$.

Consider first the case where $r_{op}=0$, so that there is no population inversion, nor lasing. In this case, the strong pump would have to be applied externally. This is illustrated in Fig. 4(a) for a resonant pump ($\Delta=0$), with $\Omega_1=2\pi\times36\times10^6 s^{-1}$. As expected, this produces the conventional Mollow-Ezekiel spectrum[12,13]. Next, we consider the case where $r_{op}=2\Gamma$, so that population inversion is present. In this case, the strong pump is the lasing mode. Figure 4(b) shows that in this case the absorption and gain features are inverted as compared to the conventional spectrum shown in Fig 4(a). Figs 4(c) and (d) show line shapes when the strong pump is detuned from resonance ($\Omega_1=2\pi\times66\times10^6 s^{-1}$, $\Delta=2\pi\times2\times10^7 s^{-1}$). Specifically, at $\omega_{ba} + \Omega' + \Delta$ ($\omega_{ba} - \Omega' + \Delta$), there is a gain (absorption) peak for $r_{op}=0$, and an absorption (gain) peak for $r_{op}=2\Gamma$. Here we define the effective Rabi frequency as $\Omega' = \sqrt{|\Omega_1|^2 + \Delta^2}$.

Using the standard notation employing Hilbert transforms, the KK relations corresponding to the susceptibility in Eqn. 10 can be expressed simply as:

$$\chi'(\delta) = \frac{1}{\pi} P.V. \left\{ \int_{-\infty}^{\infty} \frac{\chi''(\delta')}{\delta'-\delta} d\delta' \right\}; \quad \chi''(\delta) = -\frac{1}{\pi} P.V. \left\{ \int_{-\infty}^{\infty} \frac{\chi'(\delta')}{\delta'-\delta} d\delta' \right\} \quad (11)$$

Where P.V. stands for Cauchy principal value. By carrying out these integrals, it is easy to verify that these relations are indeed satisfied, with or without population inversion.

We now discuss briefly the physical interpretation of the absorption gain profiles, augmenting the comments in Ref. 13 by including the population inverted system. The strong non-resonant pump splits the energy levels through the light shift, as shown Fig. 5(a). Fig. 5(b) shows the origin of the spectral features displayed in Fig.4(c). The weak probe is maximally absorbed at the shifted transition level corresponding to $\omega_{ba} - \Omega' + \Delta$. The gain feature is explained by a three photon process: the atom absorbs two photons of the strong pump field, is excited from the lower dressed states to the higher dressed states, and emits a photon at $\omega_{ba} + \Omega' + \Delta$. Thus the probe is amplified at $\omega_{ba} + \Omega' + \Delta$. When the atoms is prepared in the excited level |b⟩, however, as shown in Fig. 5(c), the population-inverted atom emits a photon ($\omega_{ba} - \Omega' + \Delta$) which contributes to the amplification of the probe field. For the three photon process, the atom emits two photons associated with the pump field, and absorbs the probe photon at $\omega_{ba} + \Omega' + \Delta$. Therefore, the gain and loss spectra produced by the population inverted atoms are reversed as compared to the cases without population inversion.

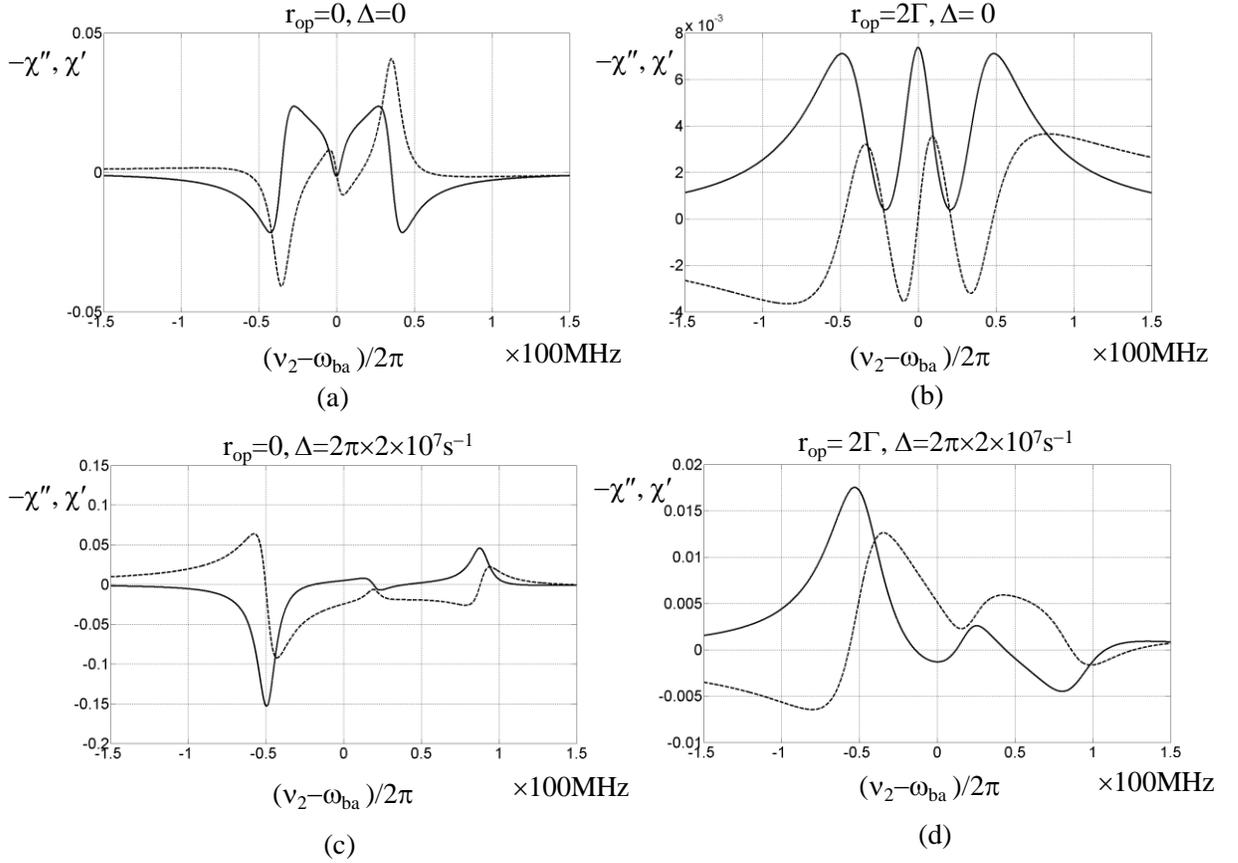

FIG. 4. Imaginary (solid lines) and real (dashed lines) parts of the susceptibility of a probe frequency $\nu_2$. For resonant pump: $\Delta=0$, (a) $r_{op}=0$ and (b) $2\Gamma$. For detuned pump: $\Delta=2\pi\times2\times10^7 s^{-1}$, (c) $r_{op}=0$ and (d) $2\Gamma$.

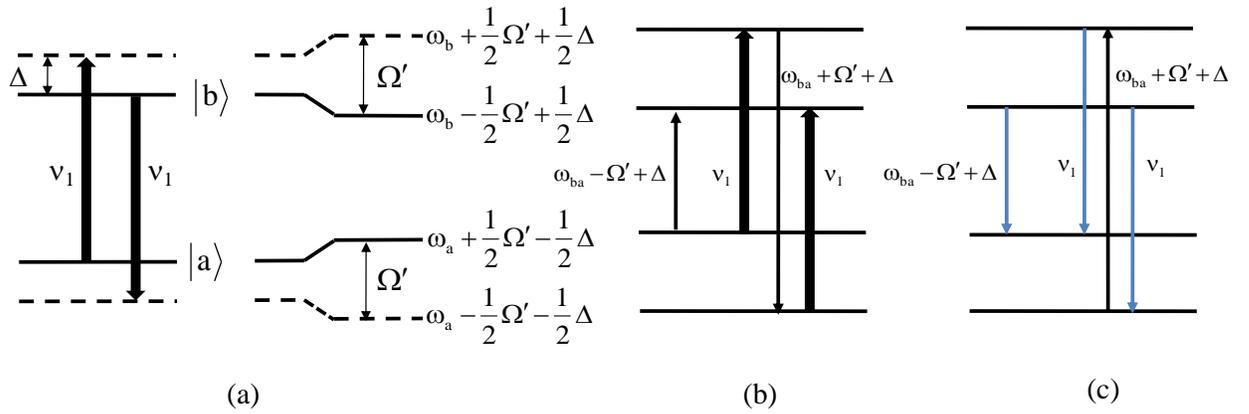

FIG.5. (a) Energy level splitting by the non-resonant pump field. (b) A probe photon of $\omega_{ba}-\Omega'+\Delta$ is absorbed by an atom, producing the absorption peak. A three photon process involving absorption of two photons from the pump and emission of one photon by the dressed atom amplifies the probe at $\omega_{ba}+\Omega'+\Delta$. (c) The atom is initially population inverted. Photons associated with the transition $\omega_{ba}-\Omega'+\Delta$ amplify the probe. In a three photon process, the atom emits two photons corresponding to $\nu_1$ and absorbs one photon of $\omega_{ba}+\Omega'+\Delta$ from the probe.

In summary, we have shown that the effective gain and dispersion profile seen by a lasing mode is non-analytical, and the KK-relations cannot be applied to relate these two profiles. We show furthermore that by applying an additional probe beam, we can produce a gain and dispersion profile that satisfy the KK-relations, and represent the only meaningful relation to the constraint imposed by causality in such a system.

This work was supported by DARPA through the slow light program under grant FA9550-07-C-0030, by AFOSR under grant FA9550-06-1-0466, and by the NSF IGERT program under grant # DGE-080-1685